\documentclass{article}

\usepackage[english]{babel}
\usepackage[pdftex]{graphicx}

\newcommand{\be}{\begin{equation}}
\newcommand{\ee}{\end{equation}}
\newcommand{\bea}{\begin{eqnarray}}
\newcommand{\eea}{\end{eqnarray}}

\newcommand{\tw}{t_{\rm w}}

\newcommand{\dd}{d}
\newcommand{\eee}{\mathrm{e}}
\newcommand{\ii}{\mathrm{i}}

\begin{document}

\title{Aging and linear response in the H\'ebraud-Lequeux model for amorphous rheology}

\author{Peter Sollich\footnote{Department of Mathematics, King's 
College London, Strand, 
London, WC2R 2LS UK, peter.sollich@kcl.ac.uk}, Julien 
Olivier\footnote{Aix Marseille Universit\'e, 
CNRS, Centrale Marseille, I2M 
UMR 7373, 
13453, Marseille, France} , Didier Bresch\footnote{Savoie Mont-Blanc 
Universit\'e, CNRS, LAMA UMR5127,
73376 Le Bourget du lac, France}}

\maketitle
\begin{abstract}
We analyse the aging dynamics of the H\'ebraud-Lequeux model, a 
self-consistent stochastic model for the evolution of local stress in an 
amorphous material. We show that the model exhibits initial-condition 
dependent freezing: the stress diffusion constant decays with time as 
$D\sim 1/t^2$ during aging so that the cumulative amount of memory that can 
be erased, which is given by the time integral of $D(t)$, is finite. 
Accordingly the shear stress relaxation function, which we determine in the 
long-time regime, only decays to a plateau and becomes progressively 
elastic as the system ages. The frequency-dependent shear modulus exhibits 
a corresponding overall decay of the dissipative part with system age, 
while the characteristic relaxation times scale linearly with age as 
expected.
\end{abstract}

\section{Introduction}
The prediction and modelling of the mechanical behaviour and flow of amorphous materials is an active area of research, reviewed recently in e.g.~\cite{RodTanVan11,CheWenJanCroYod10,Coussot07}.
Because identifying flow events -- as the analogue of dislocation motion in crystalline solids -- remains a challenge for microscopic models, {\em mesoscopic} models have been proposed as one strand of this research effort, with the goal of capturing some of the salient physics without resorting to a detailed particle-based description. Among such mesoscopic models are Shear Transformation Zone theory~\cite{FalLan98,FalLan11,Langer15}, the soft glassy rheology (SGR) model~\cite{SolLeqHebCat97,Sollich98,FieSolCat00,SolCat12}, fluidity models (see e.g.~\cite{DaCCheBonCou02}) and the H\'ebraud-Lequeux (HL) model~\cite{HebLeq98} and its variants. The latter is a stochastic model describing the evolution of the shear stress $\sigma$ of a local element of material, under the influence of an externally applied shear strain $\gamma$ and stochastic noise arising from flow events elsewhere in the material that perturb the local stress. 

The HL model, like the STZ and SGR models in their original
formulations, is a ``one-element'' model that contains coupling to
other elements of the material only via a self-consistency requirement
for the noise level. But it has also been obtained in approximate
treatments of more complicated models that explicitly represent the
spatial structure of the amorphous material under
study~\cite{BocColAjd09,ManColChaBoc11,NicMarBocBar14}. This makes it
important to understand fully the predictions of the HL model. These
have been worked out for steady shear (constant $\dot\gamma$), leading
to the flow curve $\sigma_{\rm ss}(\dot\gamma)$ giving steady state
shear stress versus shear rate~\cite{HebLeq98}. The main control parameter
$\alpha$ regulates how strongly flow events elsewhere affect a given
local element of material. For high $\alpha$ ($\alpha>1/2$), the flow
curve is linear at small shear rates, $\sigma_{\rm ss}\propto
\dot\gamma$, representing Newtonian flow. For small $\alpha$, on the
other hand, a nonzero yield stress $\sigma_{\rm ss}(\dot\gamma\to 0)$
appears: for (average) stress below this value a steady flow cannot
then be maintained. This regime can therefore be identified as the
``glassy'' one, where the amorphous material has acquired solid-like
properties, and $\alpha=1/2$ gives the location of the corresponding
glass transition~\cite{HebLeq98,OlRe1}.

The glassy regime of the HL model was analysed in the original paper~\cite{HebLeq98} only in steady states driven either by steady flow, as above, or by oscillatory strain $\gamma(t) = \gamma_0 \cos(\omega t)$ of some nonzero amplitude $\gamma_0$. In the absence of strain, however, one expects the model to display {\em aging}, i.e.\ its properties should depend on the ``waiting time'' $\tw$ since the system was prepared, also called the age of the system.
The linear stress response to an applied step strain, also called the stress relaxation function, then becomes a function $G(t,\tw)$ of both the age $\tw$ when the perturbation is applied, and the time $t$ when the stress response is measured. The goal of this paper is to establish the aging behaviour of the HL model, with a particular focus on $G(t,\tw)$ and its frequency-domain analogue.

In Sec.~\ref{sec:model} we introduce the HL model and our approach for calculating its linear response to applied strain in the aging regime. We focus on the long-time regime throughout in our analysis, which for two-time quantities like the stress relaxation function $G(t,\tw)$ means that we will consider both times large but with fixed ratio $t/\tw=1+x$, i.e.\ we take $\tw\to\infty$ at fixed $x$. 

In Sec.~\ref{sec:aging} we describe the aging behaviour of the HL model in the absence of applied strain, leaving most details of the analysis to an appendix. One of the main results will be that stochastic effects die out so quickly in the glassy regime that even after an infinite time they are not sufficient to erase memory of the initial preparation of the system.

The discussion of the linear stress response is split into two parts: in Sec.~\ref{sec:G_qualitative} we give qualitative arguments for the behaviour of $G(t,\tw)$ in the long-time regime. A precise quantitative analysis requires the use of boundary layer scaling techniques, which we discuss in Sec.~\ref{sec:G_scaling}. The use of these techniques, which were developed previously for the HL model in~\cite{Ol,OlRe1}, is the main technical contribution of this paper, making it rather distinct from approaches -- e.g.\ temporal Laplace transforms -- used to analyse aging in other mesoscopic models for amorphous rheology such as SGR~\cite{Sollich98}.

We translate our results into the frequency domain in Sec.~\ref{sec:G_omega}, which corresponds to the experimentally common technique of probing linear stress response to oscillatory strain. In Sec.~\ref{sec:numerics}, finally, we compare results from a numerical solution of the HL model to the predicted long-time asymptotics, finding good agreement. A short summary and discussion of our results is provided in Sec.~\ref{sec:summary}.

\section{The HL model}
\label{sec:model}

In dimensionless units, the H\'ebraud-Lequeux model describes the time evolution of a stress distribution $p(\sigma,t)$ as

\be
\partial_t p(\sigma,t) = -\dot\gamma \partial_\sigma p + D\partial_\sigma^2 p - h(\sigma) p +
\Gamma\delta(\sigma)
\label{HL_def}
\ee

where $h(\sigma)=\Theta(|\sigma|-1)$ and the yield rate $\Gamma(t)$ is determined by
\be 
\Gamma(t)=\int d\sigma\, h(\sigma) p(\sigma,t)\ .
\ee
as required for conservation of probability, $\int d\sigma\, p(\sigma,t) =1$.

The interested reader is
referred to~\cite{HebLeq98} for more details on the model and its interpretation. Briefly, $p(\sigma,t)$ can be viewed as the distribution of shear stresses across all local elements of an amorphous material. 
The first term on the r.h.s.\ of (\ref{HL_def}) represents elastic response $\dot\sigma=\dot\gamma$ of the stress $\sigma$ to the applied strain $\gamma$; the relevant elastic shear modulus has been scaled to unity. The third term describes yielding: once the local stress $\sigma$ exceeds a yield stress, which is again scaled to unity, a yield event occurs with unit rate. Physically this corresponds to a rearrangement of particles that relaxes the stress in an element to zero, as represented by the fourth term on the r.h.s.\ of (\ref{HL_def}). The yield rate, i.e.\ the overall at which such events take place, is $\Gamma$. 

The key self-consistent aspect of the model lies in the second term in 
(\ref{HL_def}): yield events occurring elsewhere in the material will 
perturb the local stress of an element. The assumption of the model is that 
this perturbation can be described as Gaussian noise with a stress 
diffusion constant that is proportional to the yield rate, $D(t)=\alpha 
\Gamma(t)$. (We use the conventional name ``diffusion constant'' but note 
that $D(t)$ in general depends on time.) Here $\alpha$ encodes the 
interaction strength between elements, or alternatively the ability of the 
system to propagate mechanical noise, and is the key control parameter of 
the model. The overall macroscopic stress is assumed to be given by the 
average $\langle \sigma\rangle =\int d\sigma\, \sigma p(\sigma,t)$.

For the analysis of aging in the HL model in the absence of applied strain, we will study the solution of (\ref{HL_def}) for $\dot\gamma=0$; see Sec.~\ref{sec:aging}. We then move to the linear response of the stress to small applied strains
$\gamma(t)$, for an aging system at $\alpha<1/2$. 
The simplest perturbation
scenario, from which all other linear response functions can be
calculated, is a step in $\gamma(t)$ at time $\tw$, $\gamma(t)=\gamma_0 \Theta(t-\tw)$ with small amplitude $\gamma_0$. Then for $t>\tw$ we can expand  $p$ as follows
\be
p(\sigma,t) = p_0(\sigma,t) + \gamma_0 \delta p(\sigma,t) + O(\gamma_0^2).
\label{p_expansion}
\ee
For the stress diffusion constant we have in principle similarly $D(t)=D_0(t)+\gamma_0\delta D(t)+O(\gamma_0^2)$ but it turns out that the first order perturbation $\delta D(t)$ vanishes provided we make an assumption that we will use throughout the paper, namely that the initial system preparation produces a symmetric stress distribution, $p_0(\sigma,0)=p_0(-\sigma,0)$. 

To see why $\delta D(t)=0$, note generally that (\ref{HL_def}) is invariant under the joint transformation $p(\sigma,t)\to p(-\sigma,t)$ and $\gamma(t)\to -\gamma(t)$. In the absence of strain, symmetry of the stress distribution is therefore preserved by the time evolution, i.e.\ if as assumed $p_0(\sigma,0)$ is symmetric then so is $p_0(\sigma,t)$ for all $t$. With the step strain added, the invariance of the time evolution under joint sign reversal of $\sigma$ and $\gamma_0$ then implies that
\be
p_0(\sigma,t) + \gamma_0 \delta p(\sigma,t) + O(\gamma_0^2) 
\ee
and
\be
p_0(-\sigma,t) - \gamma_0 \delta p(-\sigma,t) + O(\gamma_0^2)
\ee
are both solutions of the master equation. 
Because of the assumed symmetry of $p_0$ and uniqueness of the solution for given $p_0(\sigma,0)$ and $\gamma_0$ this tells us that $\delta p(\sigma,t)=-\delta p(-\sigma,t)$: $\delta p$ is an odd function of $\sigma$, and hence $\delta D(t) = \alpha \int d\sigma\,h(\sigma)\delta p(\sigma,t)$ vanishes.

Expanding the master equation (\ref{HL_def}) in
$\gamma_0$ and using $\delta D(t)=0$ gives then, by comparing the $O(1)$ and  $O(\gamma_0)$ contributions on
both sides,

\begin{equation}\label{eq:MainOrder}
 \partial_t p_0(\sigma,t)=
  D_0(t)\partial_{\sigma}^2p_0(\sigma,t)
 -h(\sigma)p_0(\sigma,t)+
\frac{D_0(t)}{\alpha}\delta(\sigma) 
\end{equation}
as expected, and
\be
\partial_t \delta p(\sigma,t) = D_0(t)\partial_\sigma^2 \delta p (\sigma,t) -
h(\sigma) \delta p(\sigma,t) 
\label{master_lin}
\ee

The initial condition for $\delta p$ can be obtained by integrating (\ref{HL_def}) across a small time interval around $\tw$, bearing in mind that $\dot\gamma(t)=\gamma_0\delta(t-\tw)$. This yields
\be
\delta p(\sigma,\tw)=-\partial_\sigma p_0(\sigma,\tw)
\label{delta_p_initial}
\ee

where the l.h.s.\ is to be understood as the limit of $\delta p(\sigma,t)$ for $t\to \tw$ from above.

Once we have the solution for $\delta p$, 
the quantity we are interested in is the stress relaxation function 
\begin{equation} \label{Stress-Relaxation}
G(t,\tw)=\int d\sigma\,\sigma \delta p(\sigma,t)
\end{equation}
which may be simplified to $G(t, \tw) = 2\int_0^\infty d\sigma\,\sigma \delta p(\sigma,t)$ because $\delta p$ is anti-symmetric.

The key benefit of considering a scenario where $\delta D(t)=0$, i.e.\ where the stress diffusion is unchanged in linear response, is that the linearized master equation (\ref{master_lin}) no longer has any self-consistency condition attached to it: $D_0(t)$
is determined by the unperturbed solution, rather than being
self-consistently coupled to $\delta p$.

The assumption that $p_0(\sigma,t)$ is
symmetric in $\sigma$ is also physically plausible. It applies if, for example,
one prepares the system initially at some $\alpha>1/2$, where it
reaches equilibrium in the absence of strain~\cite{HebLeq98}, and then reduces $\alpha$ to a value below $1/2$
at time $t=0$. On the other hand one would obtain a non-symmetric $p_0(\sigma,t)$ if $\alpha<1/2$ is
constant and one initially randomizes the system by pre-shear, i.e.\ by shearing at
some steady shear rate $\dot\gamma$ for a long time and then reducing the shear rate to zero at $t=0$.

\section{Aging in the absence of strain}
\label{sec:aging}

We consider in this section the behaviour of the HL model in the absence of strain. The corresponding unperturbed stress distribution $p(\sigma,t)$ evolves in time according to (\ref{eq:MainOrder})
\begin{equation}\label{eq:unperturbed_master}
 \partial_t p(\sigma,t)=
  D(t)\partial_{\sigma}^2p(\sigma,t)
 -h(\sigma)p(\sigma,t)+
\frac{D(t)}{\alpha}\delta(\sigma) 
\end{equation}
where for brevity we have dropped all ``0'' subscripts.

Following the method developed 
in \cite{Ol,OlRe1} to describe the glass transition in the HL-model, 
we expect aging at $\alpha<1/2$ to result in power law dependences on time $t$ for 
large times. The relevant ansatz for $p(\sigma,t)$ has to be split into the ``interior'' 
region $-1<\sigma<1$, where no yield events take place ($h(\sigma)=0$), and 
the ``exterior'' region $|\sigma|>1$, and the power law scaling appears in 
the boundary layers around the boundaries between these two regions.
For the interior we assume

\be
p(\sigma,t) = \sum_{k=0}^\infty t^{-k/s} Q_k(\sigma)
\ee

and for the exterior, for $\sigma>1$ and $\sigma<-1$ respectively,

\begin{eqnarray}
p(\sigma,t) &=& \sum_k t^{-k/s} R^\pm_k(t^{l/s}(\pm\sigma-1))
\end{eqnarray}
This gives for the diffusion constant
\be
D(t) = \sum_k d_k t^{-(k+l)/s}
\label{D_t}
\ee
where, with $z=
t^{l/s}(\pm\sigma-1)$,
\be
d_k = d^+_k + d^-_k, \qquad
d^\pm_k = \alpha \int_0^\infty dz \, R^\pm_k(z)
\ee
The goal is now as in \cite{Ol,OlRe1,OlRe2} to identify the integers $s$ 
and $l$ defining the scaling exponents, by inserting the above ansatz into 
the master equation (\ref{eq:unperturbed_master}) and using the boundary or ``transmission'' conditions, of 
continuity of $p$ and $\partial_\sigma p$ at $\sigma=\pm 1$. Note that in 
order to have a general framework we have not yet imposed the symmetry 
$p(\sigma,t)=p(-\sigma,t)$. We will specialize to this case shortly.

\subsection{Equations and boundary conditions for scaling functions}
In the interior, the equation of motion becomes
\be
 \sum_k -\frac{k}{s} t^{-(k+s)/s}Q_k = \sum_{k,k'} t^{-(k+k'+l)/s}
d_{k'} Q''_k + \sum_k d_k t^{-(k+l)/s} \frac{\delta(\sigma)}{\alpha}
\ee
so equality of the terms of order $t^{-m/s}$ gives
\be
-\frac{m-s}{s} Q_{m-s} = \sum_{k=0}^{m-l} d_{m-k-l} Q''_k +
d_{m-l} \frac{\delta(\sigma)}{\alpha}
\label{int}
\ee
In the exterior, one has similarly, with $z=
t^{l/s}(\pm\sigma-1)$ as before,
\be
 \sum_k t^{-(k+s)/s}\left(-\frac{k}{s} R^\pm_k
+ \frac{l}{s}z R^\pm_k{}'\right)
= \sum_{k,k'} t^{-(k+k'-l)/s}d_{k'} R^\pm_k{}'' - \sum_k t^{-k/s} R^\pm_k
\ee
and equality of the terms of order $t^{-m/s}$ gives
\be
-\frac{m-s}{s} R^\pm_{m-s}
+ \frac{l}{s}z R^{\pm'}_{m-s}
= \sum_{k=0}^{m+l} d_{m-k+l} R^\pm_k{}'' - R^\pm_m
\label{ext}
\ee
Finally, continuity of $p$ and its first derivative w.r.t.\ $\sigma$ give 
the boundary conditions
\be
Q_m(\pm 1)=R^\pm_m(0)
\label{cont}
\ee
and
\be
\pm Q_m'(\pm 1)=R^{\pm'}_{m+l}(0)
\label{contd}
\ee
 
\subsection{Results}

We defer further details of the analysis of the above equations 
to~\ref{app:aging_no_strain}. We find there that under reasonably generic 
conditions, the scaling exponents are simply $l=s=1$. From (\ref{D_t}) this 
implies in particular that the diffusion constant decays to leading order 
as $D(t) \approx d_1/t^2$. 

We also find that the leading order interior profile $Q_0(\sigma)$ remains 
undetermined by the set of equations for the scaling functions, except for 
some constraints on its derivatives at the boundary. This profile must 
therefore be dependent on initial conditions: the HL-model has ``initial 
condition-dependent'' freezing, where during aging there is not enough 
stochasticity in the stress evolution to remove all memory of the initial 
system preparation. The origin of this is the fact that the diffusion 
constant decays so quickly that $\int dt'\,D(t')$ is {\em finite}. Seeing 
as the diffusion constant is self-consistently tied to the stress 
distribution, it is then not unexpected that also $d_1$, the leading 
prefactor in $D\approx d_1/t^2$, emerges as dependent on the initial 
conditions. 

Specializing the results from~\ref{app:aging_no_strain} to symmetric 
$p(\sigma,t)$, we find in the exterior that $R_0(z)=0$ and the first 
nonzero contribution is
\be 
R_1(z) = \sqrt{d_1}/({2\alpha}) e^{-z/\sqrt{d_1}} 
\label{R1sol}
\ee
The exterior tail of $p(\sigma,t)$ is therefore a simple exponential; its 
width is determined by $d_1$, which is linked to the frozen-in profile 
$Q_0(\sigma)$ by $d_1=1/(-2\alpha Q_0''(1))^2$.

In the interior, the first subleading correction to $Q_0(\sigma)$ is
\be
Q_{1}(\sigma) = -d_{1} Q''_0(\sigma) - \frac{d_1}{\alpha}\delta(\sigma)
\label{Q1sol}
\ee
where the $\delta(\sigma)$ term just removes an equal and opposite 
contribution in $-d_1 Q''_0(\sigma)$, which arises from the fact that 
$Q_0(\sigma)$ has a kink at $\sigma=0$.

\section{Linear response to step strain: qualitative analysis}
\label{sec:G_qualitative}

In this section we give an overview of the qualitative physics that 
determines the stress relaxation function. 
As above we drop the zero subscripts on unperturbed quantities. We saw in 
Sec.~\ref{sec:aging} that
$p(\sigma,\tw)=Q_0(\sigma)+Q_1(\sigma)/\tw+\ldots$ in the interior,
and $p(\sigma,\tw)=R^{\pm}_1(\tw(\pm\sigma-1))/\tw+\ldots$ in the exterior. 
By the assumed symmetry one has
$Q_0'(\sigma=\pm1)=\mp 1/(2\alpha)$ (see (\ref{Q0constraint} in 
App.~\ref{app:aging_no_strain}), and
$R^\pm_1(z)=\sqrt{d_1}/(2\alpha)\exp(-z/\sqrt{d_1})$ decays
exponentially; $d_1$ is determined from $Q_0''(\pm 1)$.

The initial condition for the linear response problem is then $\delta
p(\sigma,\tw) = - Q_0'(\sigma)$ to leading order in the
interior, and $\delta
p(\sigma,\tw) =\pm 1/(2\alpha)\exp[-\tw(\pm\sigma-1)/\sqrt{d_1}]$ in the
exterior. This has a step discontinuity at $\sigma=0$, is of order
unity and positive for $0<\sigma<1$, and then drops to zero quickly with 
increasing $\sigma$,
on a scale $\sigma-1\sim 1/\tw$. This shows that the leading
contribution to the stress relaxation must come from the interior, with
the exterior tail giving a contribution of at most $O(1/\tw)$.

Let us  look at the evolution of $\delta p(\sigma,t)$ from the above 
initial condition under
(\ref{master_lin}), initially by approximating the unperturbed
diffusion rate $D(t)$ as constant, $D(t)\approx D(\tw)\sim 1/\tw^2$. One 
can then
calculate the eigenfunctions of the master operator in (\ref{master_lin}), 
and express the
solution as a superposition of these. The observations from this 
calculation, which we do not detail here, are
as follows:
\begin{itemize}
\item The boundary value $\delta p(1,t)$ drops quickly as $t$ increases 
from $\tw$ and then crosses over to a power law decay $\sim 1/\sqrt{t-\tw}$ 
for $t-\tw\gg 1$.
\item In the long-time limit $t-\tw=x\tw\to \infty$, so the boundary value 
vanishes. In the interior near the boundary, $\delta p(\sigma,t)$ drops 
from
  values of order unity to this vanishingly small value 
within a zone of
  size $1-\sigma \sim [D(\tw)(t-\tw)]^{1/2}$.

\item At the origin $\delta p(\sigma=0,t)$ is strictly zero 
  for any $t>\tw$, as expected from anti-symmetry. From there $\delta 
p(\sigma,t)$ rises to values of order unity
  within a distance $\sigma \sim [D(\tw)(t-\tw)]^{1/2}$.

\end{itemize}

These results, in particular the last two, are consistent with a diffusive 
``softening'' of
  the hard ``edges'' of the initial condition at $t=\tw$. They would imply
  that the stress relaxation function should decay from its initial
  value as $1-G(t,\tw) \sim [D(\tw)(t-\tw)]^{1/2}$.

In the above analysis we had assumed $t-\tw$ to be not too long, in order 
to approximate $D(t)\approx D(\tw)$ as constant. One expects the 
qualitative picture to remain the same at larger $t-\tw$, however, provided 
we replace $D(\tw)(t-\tw)$ by
  $\int_{\tw}^t dt' D(t') \sim (1/\tw-1/t)$. So we expect that in fact
  $1-G(t,\tw) \sim (1/\tw-1/t)^{1/2}$. This implies that the stress 
relaxation
  function decays incompletely, only to $1-O(\tw^{-1/2})$.
Note that if the tail of $\delta p(\sigma,t)$ has an amplitude $\sim 
1/\sqrt{t-\tw}$ as described
  above, and given that its width will be no larger than $1/\tw$, its
  contribution to $G(t,\tw)$ will be of order $(\tw\sqrt{t-\tw})^{-1}$. In 
the long-time regime this tail contribution is then indeed negligible 
compared to the effects of order $1/\sqrt{\tw}$
  from the interior.

Physically, the prediction
\be 
1-G(t,\tw) \sim (1/\tw-1/t)^{1/2}
\label{G_intuitive_form}
\ee
is consistent with the picture of the HL model
in the aging regime as being basically frozen. With $D(t)\sim 1/t^2$,
there is effectively only a finite amount $\int_{\tw}^\infty dt' D(t')\sim
1/\tw$ of stress diffusion available if we start perturbing the system at 
$\tw$, and so it
fails to relax by more than a correspondingly small amount.

\begin{figure}
\begin{center}
 \includegraphics[width=10cm,height=8cm]{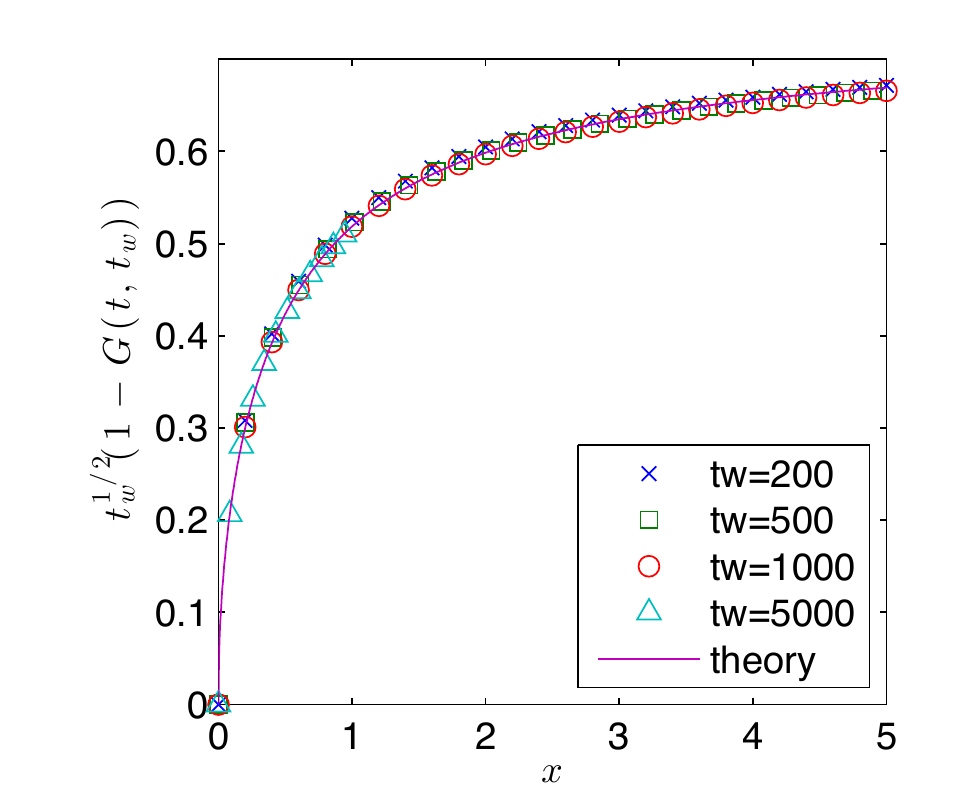}
 \caption{Scaling plot of stress relaxation function, showing the scaled 
decay of the stress relaxation function, $\tw^{1/2}(1-G(t,\tw))$, against 
$x=(t-\tw)/\tw$.
The system age $\tw$ is as shown in the legend, and $t$ varies along each 
curve; $\alpha=0.3$. All data collapse onto a master curve proportional to 
$x^{1/2}/(1+x)^{1/2}$ as expected from our analysis in the long-time 
regime. The prefactor of the 
theoretical prediction is equal to $0.733$ and is determined as explained 
in Sec.~\ref{subsec:lin_response_numerics}.
}
\label{fig:decayG}
\end{center}
\end{figure}

For a systematic analysis in the long-time regime, which we perform in the 
following section, it is useful to rewrite (\ref{G_intuitive_form}) as
\be 
\tw^{1/2}[1-G(t,\tw)]  \sim \left(\frac{x}{1+x}\right)^{1/2}
\label{eq:G_main_conjecture}
\ee 
where as previously $t=\tw(1+x)$, i.e.\ $x=(t-\tw)/\tw$. Our claim is then 
that in the limit of long times taken at constant $x$, 
$\tw^{1/2}[1-G(t,\tw)]$ becomes a function of $x$ only. We show in 
Fig.~\ref{fig:decayG} that this prediction is fully consistent with data 
obtained by direct numerical solution of the HL model, and is remarkably 
accurate even for a small waiting time $\tw=200$.

\section{Boundary layer analysis}
\label{sec:G_scaling}

\subsection{Scaling forms}
\label{subsec:scaling_setup}

Based on the intuitive discussion above, we can write down suitable scaling 
forms for the solution $\delta p(\sigma,t)$ in the long-time
regime, i.e.\ $\tw\to\infty$ with $t=\tw(1+x)$ and  $x>0$ fixed. To shorten 
calculations we fix the scaling {\em exponents} directly from the insights 
in Sec.~\ref{sec:G_qualitative}, rather than leaving them initially generic 
as we did for the unperturbed aging dynamics. The determination of the 
resulting scaling {\em functions} is the focus of this section.
 
In the {\em interior}, one expects
\be
 \delta p(\sigma,t) = 
\sum_{k\geq 0}\tw^{-k/2}\left[\delta Q_k(\sigma,x)
+B_k(\sigma\tw^{1/2},x)
+C_k((1-\sigma)\tw^{1/2},x)\right]
\ee
where the $\tw^{1/2}$ factors in $B_k$ and $C_k$ give the scale of the 
boundary
layers arising from the diffusion; the width of the boundary layers on this
scale will then grow with $x$ and eventually saturate. It is important to 
note that the size $\sim \tw^{-1/2}$
of the boundary layers, which arises from diffusive dynamics, is 
significantly larger than in the unperturbed dynamics where it is $\sim 
\tw^{-1}$.

In the {\em exterior} we expect, on the other hand,
\be
\delta p(\sigma,t) = \sum_{k\geq 0} \tw^{-k/2} \delta
R_k((\sigma-1)\tw,x).
\ee
with boundary layer size $\tw^{-1}$ inherited from the unperturbed 
solution. Note that for both the interior and exterior we have given the 
expressions as they apply for $\sigma>0$; the ones for $\sigma<0$ follow by 
anti-symmetry of $\delta p$.

We remark that the long-time limit considered here always has $t-\tw$ large 
because
we are keeping $x>0$ fixed. For time differences of order unity the scaling 
forms above do not apply: physically, they cannot describe the fast
transient that brings the value of $\delta p(1,t)$ down to $\sim
(t-\tw)^{-1/2}$. The limit of $t-\tw\gg 1$ (but of order unity) should 
nevertheless
match the $x\to 0$ limit of $t-\tw=x\tw$, and indeed we will
find a scaling with $(\tw x)^{-1/2}$ below for the leading order term
in $\delta p(1,t)$.

\subsection{Boundary and initial conditions}

We next consider the boundary and initial conditions for the boundary layer 
functions $B_k$, $C_k$, $R_k$. Antisymmetry requires the boundary condition 
at zero
\be
\delta Q_k(0,x)+B_k(0,x)=0
\label{bound0}
\ee
The boundary conditions from continuity at $\sigma=1$ are
\be
\delta Q_k(1,x)+C_k(0,x)=\delta R_k(0,x)
\label{bound1}
\ee
and from continuity of the derivative
\be
\partial_\sigma Q_m(1,x) - \partial_z C_{m+1}(0,x) = \partial_z \delta
R_{m+2}(0,x)
\label{boundd1}
\ee
The initial conditions for the interior boundary layers are that
$B_k(z,x\to 0)=C_k(z,x\to 0)=0$ for $z>0$. The initial conditions for the
$\delta Q_k$ follow from the fact that $\delta
p(\sigma,\tw)=-\partial_\sigma p_0(\sigma,\tw)$. So if
$p_0(\sigma,\tw) = \sum_k \tw^{-k} Q_k(\sigma)$, then $\delta
Q_{2k}(\sigma,0)= - \partial_\sigma Q_k(\sigma)$ while $\delta
Q_{2k+1}(\sigma,0) = 0$. The initial conditions for the $\delta R_k$
are more subtle but fortunately are not needed because these functions are
adiabatically slaved to the interior as discussed below.

For the following analysis, in order to be able to use expansions in 
$\tw^{-1/2}$ rather than $\tw^{-1}$ throughout, we will also write the 
unperturbed diffusion rate as 
\be
D(t)=\sum_{l\geq 2}
\tilde{d}_l t^{-l/2-1}=\sum_{l\geq 2} \tilde{d}_l
\tw^{-l/2-1}(x+1)^{-l/2-1}
\ee
with $\tilde{d}_{2k}=d_k$ and
$\tilde{d}_{2k+1}=0$.

\subsection{Determination of scaling functions}

We can now write down the equations for the various scaling functions
that follow from the linearized master equation (\ref{master_lin}).  
We assume that these functions decay quickly
(faster than power law) when their first argument $z$ becomes large. 
In the interior $0<\sigma<1$ 
the boundary layers then do not contribute, so
\be
 \sum_{k\geq 0}\tw^{-k/2-1}\partial_x\delta Q_k(\sigma,x)
= \sum_{l\geq 2} \tilde{d}_l \tw^{-l/2-1} (x+1)^{-l/2-1}
\sum_{k\geq 0}\tw^{-k/2}\partial_\sigma^2\delta Q_k(\sigma,x)
\ee
and therefore 
\be
\partial_x\delta Q_m(\sigma,x)
= \sum_{l=2}^m \tilde{d}_l 
(x+1)^{-l/2-1} \partial_\sigma^2\delta Q_{m-l}(\sigma,x)
\ee
Because only even $l$ contribute to the sum, one sees that the even
and odd $\delta Q_m$ decouple, and given that only even ones feature
in the initial condition at $x=0$, the odd ones will vanish at all
$x$.  The leading order term is $\partial_x \delta Q_0 = 0$, so that
$\delta Q_0(\sigma,x)=\delta Q_0(\sigma,0)$ is independent of the
rescaled time difference $x$. Note that each higher-order $\delta Q_m$
is determined simply by integrating over $x$ a function of $\sigma$
and $x$ that is known from the lower orders. In particular one is {\em
  not} solving a diffusion equation here, so one does not need separate
boundary conditions for the $\delta Q_m$ themselves. This is because
the continuity at the boundaries is handled by the boundary layer
functions $B_k$ and $C_k$.

 In the interior near $\sigma=0$, i.e.\ for fixed $z=\sigma \tw^{1/2}$
of order unity one gets from (\ref{master_lin})
\bea
 &&\sum_{k\geq 0}\tw^{-k/2-1}\left[
\sum_{n\geq 0}\frac{1}{n!}z^n \tw^{-n/2} \partial_\sigma^n
\partial_x \delta Q_k(0,x)
+ \partial_x B_k(z,x)\right]
\\
 &=& 
\sum_{l\geq 2} \tilde{d}_l \tw^{-l/2-1} (x+1)^{-l/2-1}
\sum_{k\geq 0}\tw^{-k/2}\left[
\sum_{n\geq 0}\frac{1}{n!}z^n \tw^{-n/2} \partial_\sigma^{n+2}\delta 
Q_k(0,x)
+ \tw \partial_z^2 B_k(z,x)\right]
\nonumber
\eea
and so
\bea
 &&
\!\!\!\!\!\!\!\!\!\!\!\!\!\!\!\!\!\!\!\!\!\!\!\!
\sum_{k=0}^m
\frac{z^{m-k}}{(m-k)!} \partial_\sigma^{m-k}
\partial_x \delta Q_k(0,x)
+ \partial_x B_m(z,x) \ =
\\
 &=& 
\sum_{l\geq 2}^{k+l} \sum_{k\geq 0}^{\leq m}
\tilde{d}_l (x+1)^{-l/2-1} \frac{z^{m-k-l}}{(m-k-l)!}
\partial_\sigma^{m-k-l+2}\delta Q_k(0,x)
\nonumber\\
&&{}+
\sum_{k=0}^m \tilde{d}_{m+2-k}(x+1)^{-(m-k)/2-2}
\partial_z^2 B_k(z,x)
\nonumber
\eea
The first term on the left hand side is, from the equation for
$\delta Q_m$,
\bea
&&\sum_{k=0}^m
\sum_{l=2}^k \tilde{d}_l 
(x+1)^{-l/2-1} 
\frac{z^{m-k}}{(m-k)!}
\partial_\sigma^{m-k+2}\delta Q_{k-l}(0,x)
\eea
or with $k=k'+l$
\bea
&&\sum_{l=2}^m
\sum_{k'=0}^{m-l} \tilde{d}_l 
(x+1)^{-l/2-1} 
\frac{z^{m-k'-l}}{(m-k'-l)!}
\partial_\sigma^{m-k'-l+2}\delta Q_{k'}(0,x)
\eea
which is identical to the first term on the right. So we end up with
\bea
\partial_x B_m(z,x)
&=& 
\sum_{k=0}^m \tilde{d}_{m+2-k} (x+1)^{-(m-k)/2-2}
\partial_z^2 B_k(z,x)
\eea
Again the odd and even terms decouple, and as for the odd ones the
boundary condition (\ref{bound0}) and the initial condition are zero,
these functions will vanish. The even ones can be determined
recursively, starting from $B_0$ which obeys
\bea
\partial_x B_0(z,x)
&=& 
\tilde{d}_{2} (x+1)^{-2} \partial_z^2 B_0(z,x)
\eea
and has boundary condition $B_0(0,x) = \delta Q_0(0,x) = \delta
Q_0(0,0)$ and initial condition $B_0(z,x\to 0)=0$ for $z>0$. The
solution of this is
\bea\label{eq:B0}
B_0(z,x)=\delta Q_0(0,0)\left\{-1+\mbox{erf}(z/[2(\tilde{d}_2
x/(1+x))^{1/2}])\right\}
\label{B0_result}
\eea

Next we look at $\sigma$ close to 1, i.e.\ $z=(1-\sigma)\tw^{1/2}$
finite. Then one derives from the master equation, exactly as for
$\sigma$ close to zero, for the $C_m$ the equations
\bea
\partial_x C_m(z,x)
&=& 
\sum_{k=0}^m \tilde{d}_{m+2-k} (x+1)^{-(m-k)/2-2}
\partial_z^2 C_k(z,x)
\eea
The difference to the $B_m$ is that to get the relevant boundary
condition, we also need the solution on the other side of $\sigma=1$,
i.e.\ the $\delta R_k$. For these one gets from the master
equation and with $z=(\sigma-1)\tw$
\bea
 \sum_{k\geq 0} \tw^{-k/2} \tw^{-1}\partial_x \delta
R_k(z,x) &=& 
\sum_{l\geq 2} \tilde{d}_l \tw^{-l/2-1}(x+1)^{-l/2-1}
\sum_{k\geq 0} \tw^{-k/2}\tw^2 \partial_z^2 \delta
R_k(z,x)
\nonumber
\\
&&{}
-\sum_{k\geq 0} \tw^{-k/2} \delta R_k(z,x)
\eea
and so
\bea
 \delta R_m(z,x) &=&
- \partial_x \delta R_{m-2}(z,x) + 
\sum_{l\geq 2} \tilde{d}_l (x+1)^{-l/2-1}
\partial_z^2 \delta R_{m+2-l}(z,x)
\eea
The first three of these equations are
\bea
 \delta R_0(z,x) &=& \tilde{d}_2(x+1)^{-2} \partial_z^2 \delta R_0(z,x)\\
 \delta R_1(z,x) &=& \tilde{d}_2(x+1)^{-2} \partial_z^2 \delta R_1(z,x)\\
 \delta R_2(z,x) &=& -\partial_x \delta R_0(z,x) +
\tilde{d}_2(x+1)^{-2} \partial_z^2 \delta R_2(z,x) +
\tilde{d}_4(x+1)^{-3} \partial_z^2 \delta R_0(z,x)
\eea
The time derivative ($\partial_x$) terms are always subleading, which
is consistent with the idea that the tails of $\delta p$ are evolving 
essentially
adiabatically. One can then proceed to solve order by order. The first 
equation is solved by
\bea
\delta R_0(z,x) &=& r_0(x) e^{-z(x+1)/\tilde{d}_2^{1/2}}
\eea
But the derivative boundary condition (\ref{boundd1}) tells us that
$\partial_z R_0(0,x)=0$, which implies $r_0(x)=0$, i.e.\ $\delta
R_0(z,x)=0$. Then the boundary condition (\ref{bound1}) gives $\delta
Q_0(1,x) + C_0(0,x)=0$, hence $C_0(0,x)=-\delta Q_0(1,x)=-\delta
Q_0(1,0)$. We can now solve for $C_0$, which by analogy with $B_0$
becomes
\bea\label{eq:C0}
C_0(z,x)=\delta Q_0(1,0)\left\{-1+\mbox{erf}(z/[2(\tilde{d}_2
x/(1+x))^{1/2}])\right\}
\label{C0_soln}
\eea
The functional form of the dependence on $z$ is consistent with the 
qualitative discussion in Sec.~\ref{sec:G_qualitative}, reflecting the fact 
that the stress dynamics for $\sigma<1$ is diffusive, with effectively an 
absorbing boundary at $\sigma=1$.
At the next order, the solution for $\delta R_1$ has the form
\bea
\delta R_1(z,x) &=& r_1(x) e^{-z(x+1)/\tilde{d}_2^{1/2}}
\eea
The derivative boundary condition is
\bea
-\partial_z \delta R_1(0,x) &=& \partial_z C_0(0,x)
\ = \
\delta Q_0(1,0)\frac{2}{\sqrt{\pi}[2(\tilde{d}_2
x/(1+x))^{1/2}]}
\eea
This must equal $r_1(x)(x+1)\tilde{d}_2^{-1/2}$, which fixes $r_1(x)$
and so

\bea\label{eq:dR1}
\delta R_1(z,x) &=&
\frac{\delta Q_0(1,0)}{\sqrt{\pi x(1+x)}}
\,e^{-z(x+1)/\tilde{d}_2^{1/2}}
\eea
For small $x$ the amplitude of this scales as $x^{-1/2}$, and so the
leading tail has overall amplitude
$\tw^{-1/2}x^{-1/2}=(t-\tw)^{-1/2}$, as argued above. The actual
initial condition at $t=\tw$ (which is not directly accessible by our 
long-time scaling
with $x>0$ and then $\tw\to\infty$) can be
estimated by extrapolating to $t-\tw$ of order unity, i.e.\ $x\sim
1/\tw$, and one then finds as expected that the tail amplitude is initially
of order unity.

As at the previous order, knowledge of $\delta R_1$ determines the
boundary condition (\ref{bound1}) $C_1(0,x)=\delta R_1(0,x)$ for
$C_1$. The latter can then be found from
\bea
\partial_x C_1(z,x)
&=& 
\tilde{d}_{2} (x+1)^{-2} \partial_z^2 C_1(z,x)
\eea
where the explicit result involves typical diffusive quadratic exponentials
of $z$, and $z\,\mbox{erfc}(\cdots z)$. This then gives the required
boundary condition on $\delta R_2$, and one can continue to solve
iteratively in this way.

\subsection{Evaluation of the stress relaxation function}

Using the scaling forms for $\delta p$ discussed in 
Sec.~\ref{subsec:scaling_setup}, the stress relaxation function 
$G(t,\tw)=2\int_0^\infty d\sigma\,\sigma\,p(\sigma,t)$ can be written in 
the form
\bea
 G(t,\tw) &=&
\sum_{k\geq 0}\tw^{-k/2}\Biggl[
2\int_0^1d\sigma\,\sigma\,\delta Q_k(\sigma,x)
+2\tw^{-1}\int_0^\infty dz\,z\,B_k(z,x)
\\
 &&{}
+2\tw^{-1/2}\int_0^\infty dz\,(1-z\tw^{-1/2})C_k(z,x)\Biggr]
\nonumber
\\
 &&{}
 + \sum_{k\geq 0} \tw^{-k/2} \tw^{-1} \int_0^\infty dz\,
(1+z\tw^{-1}) \delta R_k(z,x)
\nonumber
\eea
From the fact that $\delta Q_0(\sigma,x)=\delta
Q_0(\sigma,0)=-\partial_\sigma Q_0(\sigma)$ and that $Q_0(\sigma)$
must integrate to one between $\sigma=-1$ and $\sigma=1$, the first
term in square brackets is unity. The only term that contributes to
the leading order decay from this initial value is $C_0$, giving to
this leading order
\bea
\tw^{1/2}[1-G(t,\tw)]&=& -2\int_0^\infty dz\,C_0(z,x)
\eea
One can insert the explicit solution (\ref{C0_soln}) to get
\bea
\tw^{1/2}[1-G(t,\tw)]& = & 4\,\delta Q_0(1,0)
\biggl(\frac{\tilde{d}_2}{\pi}\biggr)^{1/2} 
\left(\frac{x}{1+x}\right)^{1/2}
\label{eq:G_t_tw_final}
\label{eq:Gasymp}
\eea
This depends on the frozen-in part of the aging solution via $\delta
Q_0(1,0) = - \partial_\sigma Q_0(1)$ and via $\tilde{d}_2=d_1$. We have 
therefore established the result (\ref{eq:G_main_conjecture}) obtained 
earlier from qualitative arguments, and identified the relevant, 
initial-condition dependent, prefactors.

\section{Linear response to oscillatory strain}
\label{sec:G_omega}

We now turn to the study of oscillatory shear. From the relaxation function 
$G$ one can obtain information on 
the response to oscillatory strain, $\gamma(t) \propto \eee^{\ii\omega t}$ 
(with the real part giving the physical strain). The stress response is of 
the form $\langle \sigma\rangle(t) = G^*(\omega)\gamma(t)$. In 
time-translation invariant systems, where the stress relaxation function 
$G(t,\tw)$ depends only on $t-\tw$, the complex shear modulus or 
``viscoelastic spectrum'' $G^*(\omega)$ is then proportional to the Fourier 
transform of $G$.

For aging fluids the situation is more subtle because time translation 
invariance is lost (see~\cite{FieSolCat00}). The most 
general description of the complex modulus is then as a function of the 
time $\tw$ when the strain is switched on, of the frequency $\omega$ and of 
the time $t$ when the stress is measured:
\begin{equation}\label{eq:fullGstar} 
G^*(\omega,t,\tw)=G(t,\tw)\eee^{-\ii\omega(t-t_w)}+\ii\omega\int_{\tw}
^tdt' 
G(t,t')\eee^{-\ii\omega(t-t')}.
\end{equation}
However in the long-time regime 
function $G^*$ may be close to the ``forward spectrum'' defined by
\begin{equation}\label{eq:fwdGstar}
 G^*(\omega,t)=\ii\omega\int_t^{\infty}dt'G(t',t)\eee^{-\ii\omega(t'-t)}.
\end{equation}
which is calculated as if the strain was applied from $t$ into the future. 
Physically, the conditions for this approximation to hold 
are~\cite{FieSolCat00} that $\omega \ll 1$ (relaxation timescales are large 
for long time so we need to look at small frequencies), $\omega \tw\gg 1$ 
(strain starts sufficiently late after initial preparation to make 
transient effects small) and $\omega(t-\tw)\gg 1$ (many cycles of strain 
are performed before a measurement is taken).
We show in Appendix~\ref{app:G_omega} that these expectations are indeed 
correct for the HL-model, by comparing the explicit expressions for 
$G^*(\omega,t,\tw)$ and $G^*(\omega,t)$ in the long-time limit.

We can thus focus on the forward spectrum.
To evaluate this in the long-time limit, we insert 
$G(t',t)=1-c(t^{-1/2}-t'^{-1/2})$ from (\ref{G_intuitive_form}) into
(\ref{eq:fwdGstar}), with
$c=4\,\delta Q_0(1,0)
({\tilde{d}_2}/{\pi})^{1/2}$ the prefactor
identified in (\ref{eq:G_t_tw_final}). 
Changing integration variable to $w'=\omega(t'-t)$ then gives
\begin{equation}
G^*(\omega,t) = 1- \frac{c}{\sqrt{t}} \,\ii \int_0^{\infty} dw' 
\sqrt{\frac{w'}{w+w'}} \eee^{-\ii w'}
\label{forward_spectrum_long_time}
\end{equation}
with the scaled frequency $w=\omega t$. The integral can be expressed in 
terms of Hankel functions (see~\ref{app:G_omega}) but in fact we only need 
to know its behaviour for large $w$ as that is the regime where the forward 
spectrum is physically meaningful. Here, anticipating that the integral is 
dominated by $w'$ of order unity, we can approximate $[w'/(w+w')]^{1/2} 
\approx (w'/w)^{1/2}$. The integral then becomes proportional to a Gamma 
function so that
\be
G^*(\omega,t) \approx  1- \frac{c}{\sqrt{w t}} \frac{\ii 
\sqrt{\pi}}{2\ii^{3/2}}
= 1 - (1-\ii) \frac{c\pi}{\sqrt{8 w t}}
\label{G_omega_final}
\ee
or for the real and imaginary part
\be
G'(\omega,t) \approx 1-\frac{c\pi}{\sqrt{8 w t}}
\ee
\be
G''(\omega,t) \approx \frac{c\pi}{\sqrt{8 w t}}
\ee
In these expressions one can of course equivalently write $\sqrt{w t} = 
\sqrt{\omega} t$. 
The former version makes it clearer that $\sqrt{t}[1-G^*(\omega,t)]$
becomes a function of the scaling variable $w$ for large $t$. This 
conclusion is physically sensible, and mirrors 
the fact that $\sqrt{\tw}[1-G(t,\tw)]$ becomes a function of 
$x=(t-\tw)/\tw$ for long times: for conventional aging, where the 
amplitude of the decay of $G(t,\tw)$ is independent of $\tw$, one would 
expect $G^*(\omega,t)$ itself to become a function of $w$. Here one needs 
to multiply $1-G^*$ by $\sqrt{t}$ to compensate for the $1/\sqrt{t}$ 
dependence of the decay amplitude.

\section{Comparison with numerical results}
\label{sec:numerics}

We have already provided in Fig.~\ref{fig:decayG} numerical data that 
clearly supports our main prediction (\ref{eq:G_main_conjecture}) for the 
decay of the stress relaxation function. In this section we give a more 
detailed comparison with numerics, and in particular we validate the 
boundary layer scaling forms assumed in our analysis.

\begin{figure}
\begin{center}
\includegraphics[width=10cm,height=8cm]{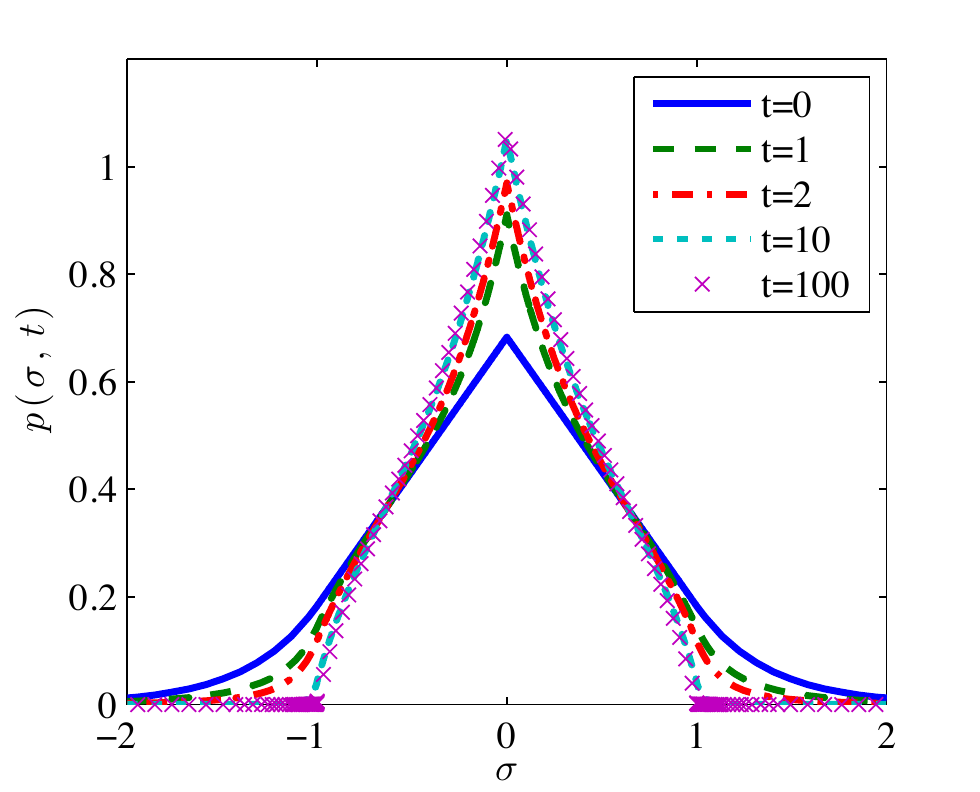}
\includegraphics[width=10cm,height=8cm]{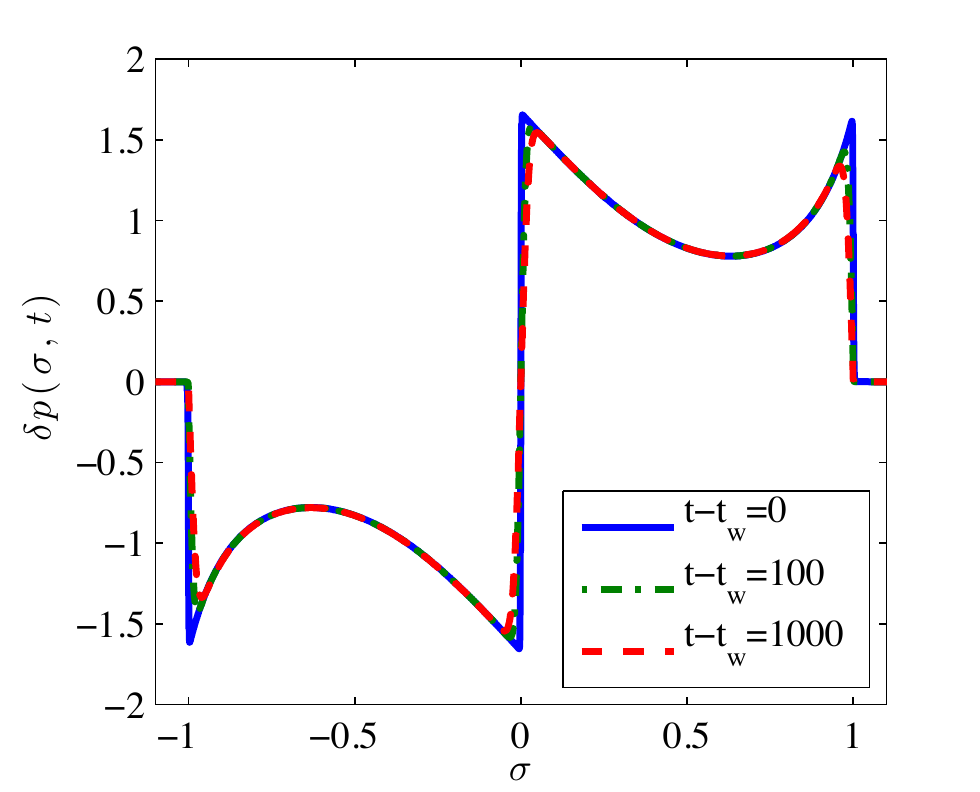}
\caption{
(a) Snapshots of numerical solution $p(\sigma,t)$ for times $t=0$, 1, 2, 
10, 100. Note the progressive freezing in the interior range $-1<\sigma<1$ 
and the shrinking boundary layer in the exterior.
(b) Snapshots of linear perturbation $\delta p(\sigma,t)$ due to step 
strain at time $\tw=200$, for $t-\tw=0$, $100$, $1000$. Here there 
is an additional boundary layer around $\sigma=0$.
\label{fig:p_overview}
}
\end{center}
\end{figure}

We focus on a setting broadly representative of a ``crunch'', i.e.\ sudden 
change in density of the material at time zero by compression. The initial 
stress distribution $p_0(\sigma,0)$ is chosen as the  
stationary solution of the HL-model~\cite{HebLeq98} for the value 
$\alpha=1$, well outside the glass phase. The crunch at time zero is 
assumed to bring the system into the glassy regime at some $\alpha<1/2$ 
that then stays constant in time; we assume in particular $\alpha=0.3$. We 
then compute the numerical solution $p_0(\sigma,t)$ of (\ref{eq:MainOrder}) 
for this setting. For a series of waiting times $\tw$ we also obtain 
$\delta p(\sigma,t)$ by solving (\ref{master_lin}) with initial condition 
(\ref{delta_p_initial}), and finally compute $G(t,\tw)$ from 
(\ref{Stress-Relaxation}). We give an overview of the numerical results for 
$p(\sigma,t)$ and $\delta p(\sigma,t)$ in Fig.~\ref{fig:p_overview}.

\subsection{Numerical methods}

The numerical implementation of the above programme is not trivial: we need 
to use a discrete grid or ``mesh'' of $\sigma$-values; but as the solution 
of the problem 
develops boundary layers whose size decreases in time, also the mesh size 
needs to decrease to obtain accurate results.

Calculations were therefore performed with a combination of a 
(one-dimensional) finite-volume discretization of the PDE and a mesh 
refinement 
algorithm. Mesh refinement is based on a standard curvature estimate 
(see~\cite{HuRu} for instance), taking into account that the diffusion 
coefficient of the linearized master equation (\ref{master_lin}) decays 
with time. While we cannot give quantitative error bounds for the accuracy 
of this approach, it certainly does refine at the locations where boundary 
layers are expected to appear. Small numerical artefacts are occasionally 
visible in the results, but usually only right before a refinement is made.

We have also carried out calculations with a constant mesh (using the most 
refined mesh obtained in 
a run with refinement but otherwise identical  parameters), in order to 
separate the effects of mesh refinement from errors resulting from the 
discretization itself. The results of the two approaches are consistent 
with each other.
\subsection{Aging without strain}

\begin{figure}
 \begin{center}
 \includegraphics[width=10cm,height=8cm]{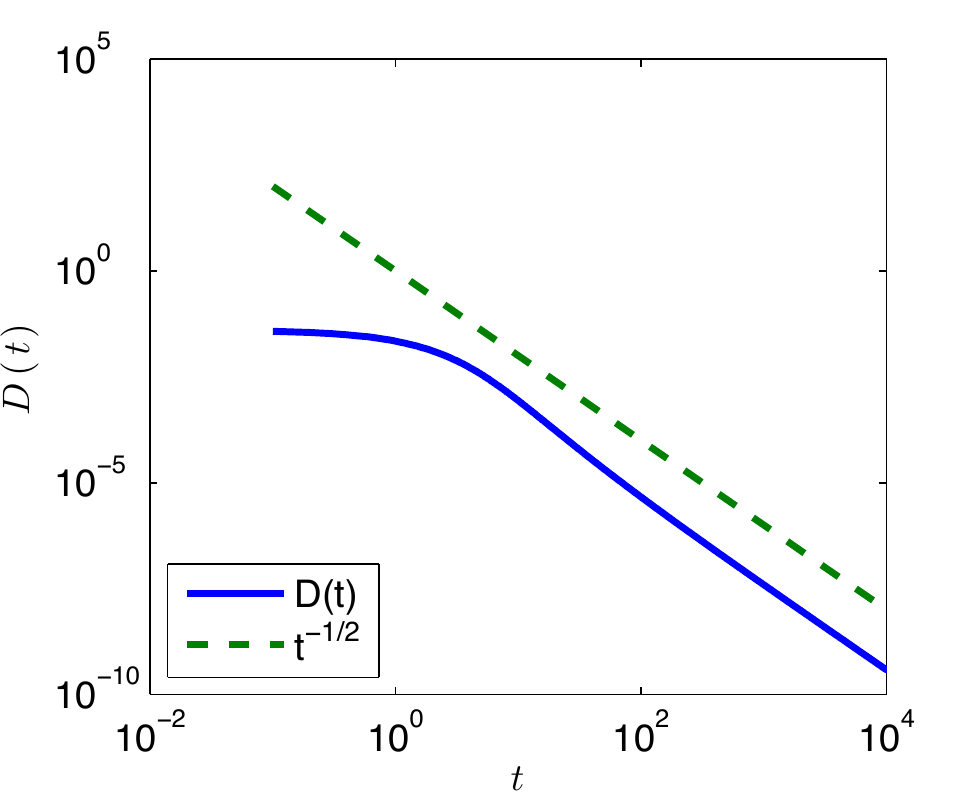}
  \caption{Log-log plot showing decay of diffusion constant $D(t)$ with 
time $t$for aging in the absence of strain. The asymptotic decay is $D\sim 
1/t^2$ for long times, as indicated by the dashed line.}
  \label{fig:diffusion}
 \end{center}
\end{figure}

We first validate the scaling ansatz for $p(\sigma,t)$ around $\sigma=\pm 
1$ and the time evolution of the diffusion constant. Starting with the 
latter, Fig.~\ref{fig:diffusion} shows that the theoretical expectation 
$D(t)\sim 1/t^2$ is well satisfied for long times, though reasonably large 
$t\approx 10^2$ are needed to clearly see this asymptotic regime 
Using the value of 
$D(t)$ at the largest time in our numerics, we estimate the asymptotic 
prefactor in $D(t)\approx \tilde{d}_2/t^2$ as 
Convergence to a constant for large $t$  seems clear but as we can see, 
this convergence is, at least in this test case, very late. One must wait 
at least before one can consider the asymptotic regime to be attained. By 
using the last computed value to approximate $d_2$, one 
\begin{equation}
\tilde{d}_2\approx 0.03804\label{eq:d2estimate}
\end{equation}
For later use we note 
the corresponding value
$1/\tilde{d}_2^{1/2}\approx 5.127$.

\begin{figure}
\begin{center}
 \includegraphics[width=10cm,height=8cm]{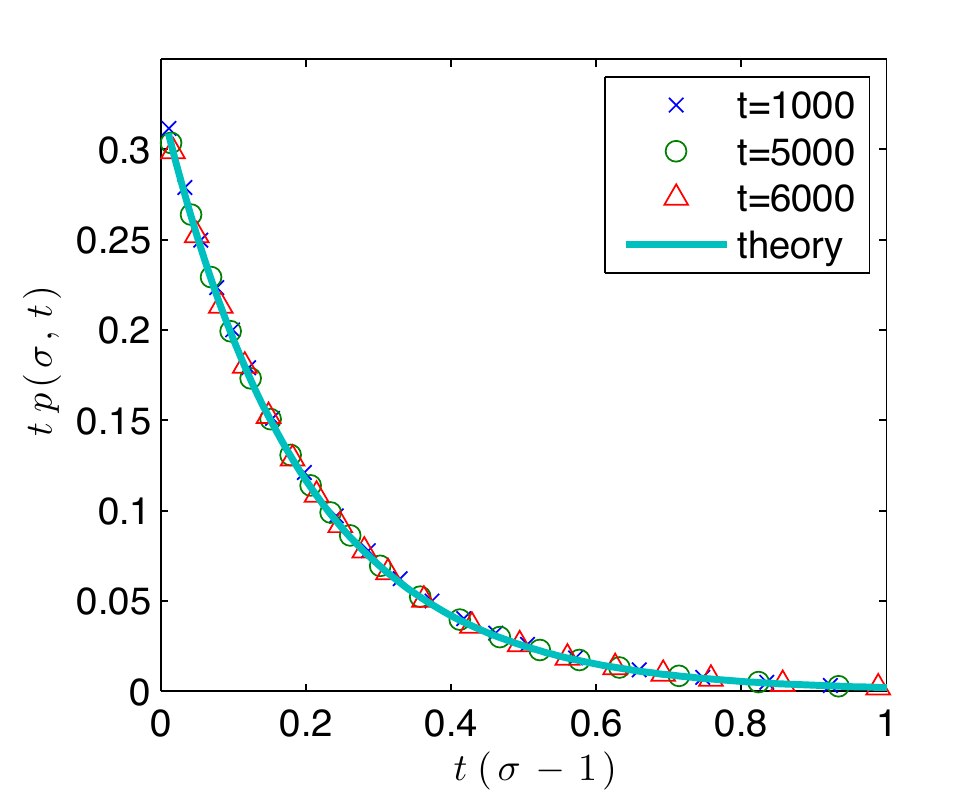}
 \caption{Scaling plot of exterior tail behaviour for aging without 
strain, 
showing $tp(\sigma,t)$ versus $z=t(\sigma-1)$ for $t=1000, 5000, 6000$
The bold line gives the theoretical prediction $[\tilde 
d_2^{1/2}/(2\alpha)]\exp(-z/\tilde d_2^{1/2})$ with $\tilde d_2$ 
given by (\ref{eq:d2estimate}).}
 \label{fig:unperturbed_tail}
 \end{center}
\end{figure}

We can verify our theoretical approach in more detail by studying the 
tail behaviour of $p(\sigma,t)$ around $\sigma=1$. According 
to Eq. (\ref{eq:dR1}) 
this is
\be
p(t,\sigma)\approx \frac{\tilde 
d_2^{1/2}}{2\alpha}\,\frac{1}{t}e^{-t(\sigma-1)/\tilde d_2^{1/2}}
\ee
to leading order for large $t$. This means 
that plots of $tp(t,\sigma)$ against 
$z=t(\sigma-1)$ for different $t$ should collapse onto the master curve 
$[\tilde d_2^{1/2}/(2\alpha)]\exp(-z/\tilde d_2^{1/2})$. We demonstrate 
this in Figure \ref{fig:unperturbed_tail}, using the value for $1/\tilde 
d_2^{1/2}$ estimated from the diffusion constant data. Consistency with 
the 
theory can also be checked in the other direction: a plot of 
$\ln[tp(\sigma,t)]$ against $z$ should be a straight line of slope 
$1/\tilde d_2^{1/2}$. Performing such a fit gives 
$1/d_2^{1/2}\approx 5.125$ in good agreement with the value 
(\ref{eq:d2estimate}) from the 
diffusion data.

\subsection{Linear response}
\label{subsec:lin_response_numerics}

\begin{figure}
\begin{center}
 \includegraphics[width=10cm,height=8cm]{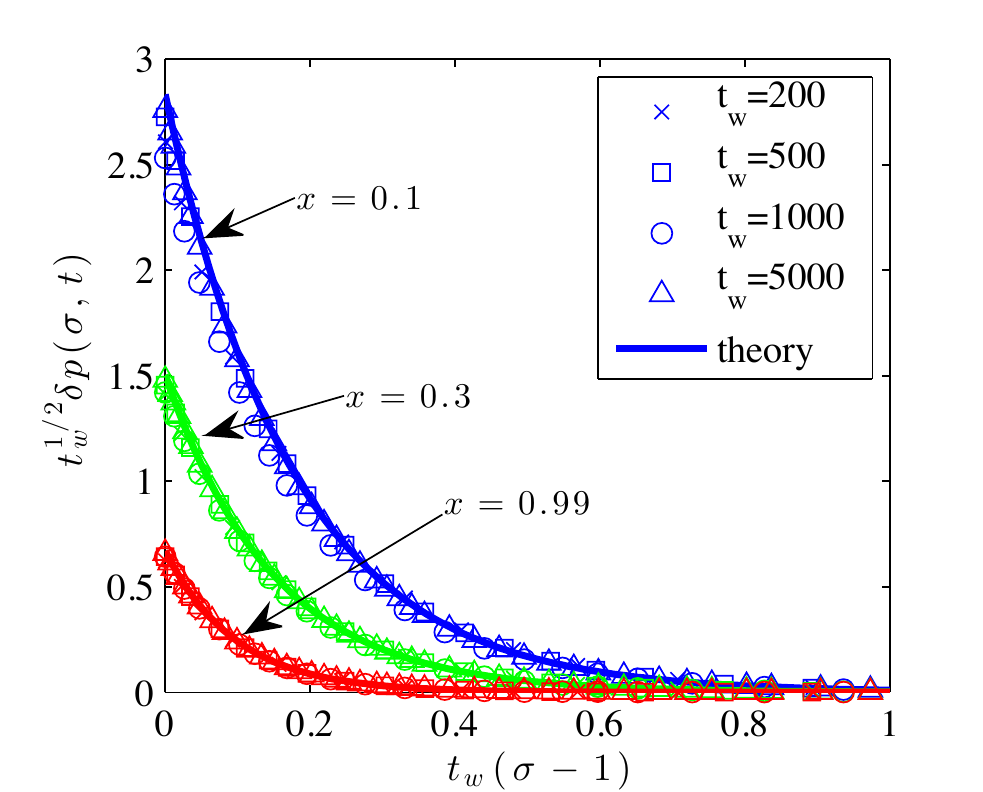}
 \caption{Decay of the linear perturbation solution $\delta 
p(\sigma,t)$ in the exterior boundary layer. Shown is
$\tw^{1/2}\delta p(\sigma,t)$ versus $\tw(\sigma-1)$ for different 
$x=(t-\tw)/\tw)$ -- as indicated by the arrows and respective colours -- 
and different $\tw$ as shown by the symbols.
Solid lines give the prediction from the 
scaling theory for $\tw\to\infty$.}
 \label{fig:perturbed_tail}
 \end{center}
\end{figure}

We next consider the behaviour of the linear response $\delta 
p(\sigma,t)$. 
We begin with the exterior tail ($\sigma>1$.) To leading order for large 
times, our scaling theory predicts for this
\begin{equation}
 \delta p(\sigma,t)\approx \frac{1}{\sqrt{\tw}} \frac{\delta 
Q_0(1,0)}{\sqrt{\pi x(1+x)}}
\,e^{-\tw(\sigma-1)(x+1)/\tilde{d}_2^{1/2}}
\label{delta_p_leading_tail}
\end{equation}
To verify this, we plot $\tw^{-1/2}\delta p(\sigma,t)$ versus 
$\tw(\sigma-1)$ in Fig.~\ref{fig:perturbed_tail} for several different 
values of $x$. The data are clearly consistent with the scaling prediction, 
with deviations that are surprisingly small even when $t-\tw$ is not very 
large (i.e.\ for small $x$ and $\tw$; for $x=0.1$ and $\tw=200$ one has 
$t-\tw=20$).
For the prefactor of the theoretical prediction we need to  estimate the 
prefactor 
$\delta Q_0(1,0)=-\partial_{\sigma} Q_0(\sigma=1)$. We do this by taking 
the numerical 
derivative of $p_0$, at the largest time available in our numerics since
$p_0(\sigma,t)=Q_0(\sigma)+O(1/t)$. Note that the resulting estimate of
$\delta Q_0(1,0)\approx 1.666$ 
is also used in determining the prefactor $4\delta Q_0(1,0)(\tilde 
d_2/\pi)^{1/2}$ of the theoretical prediction shown in 
Fig.~\ref{fig:decayG}.

\begin{figure}
\begin{center}
 \includegraphics[width=10cm,height=8cm]{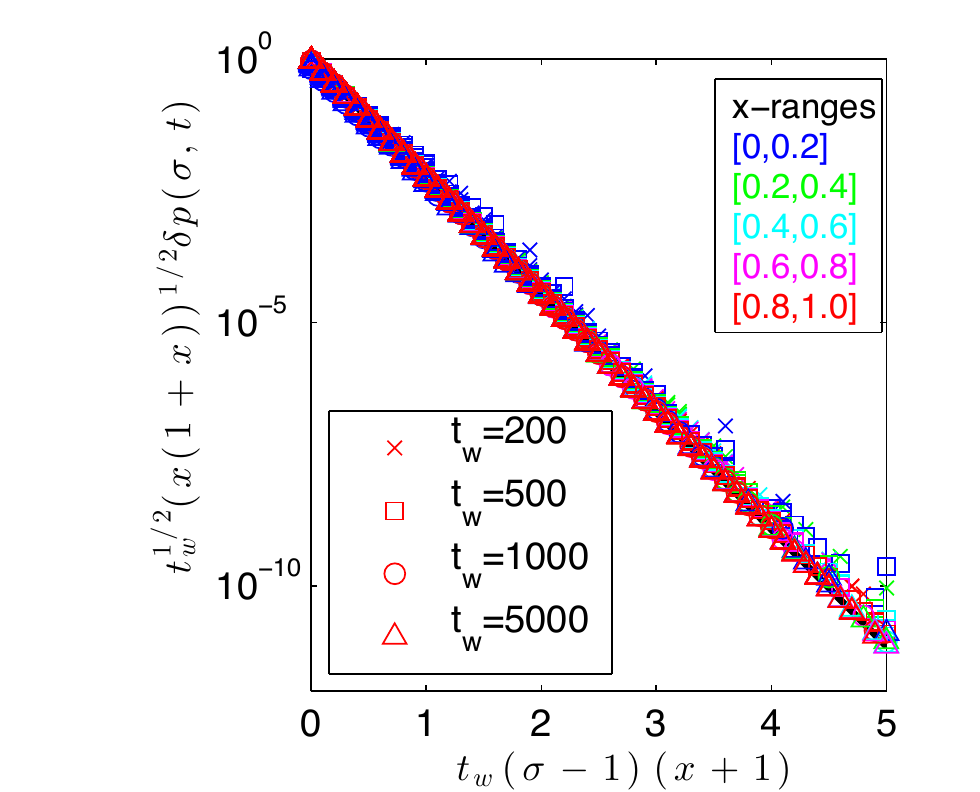}
 \caption{
Master curve for tail behaviour of the linear perturbation solution 
$\delta 
p(\sigma,t)$. Shown is
$\tw^{1/2}[x(1+x)]^{1/2}\delta p(\sigma,t)$ versus $\tw(\sigma-1)(x+1)$ 
for a range of different $x=(t-\tw)/\tw$ and $\tw$. The value of $\tw$ is 
indicated by the symbol used, as in Fig.~\ref{fig:perturbed_tail}. Colours 
identify the range of $x$ as shown in the legend. Black solid line: 
prediction from the scaling theory.
}
 \label{fig:fitting_perturbed_tail}
 \end{center}
\end{figure}
To check the results for a broader range of $x$, we plot in 
Fig.~\ref{fig:fitting_perturbed_tail}
$[\pi 
x(1+x)]^{1/2}\tw^{1/2}\delta p(\sigma,t)$ on a logarithmic axis against 
$t_w(\sigma-1)(x+1)=t(\sigma-1)$. This plot should be a straight line from 
(\ref{delta_p_leading_tail}), and again the data closely follow this 
prediction. Deviations become visible only for small $x$ and $\tw$, and 
primarily in the regime where the scaled $\delta p$ is already very small.
As before one can check consistency in the reverse direction also, by 
fitting the slope of the straight line in 
Fig.~\ref{fig:fitting_perturbed_tail}. This produces the estimate 
$1/d_2^{1/2}\approx 5.132$, again in good agreement with the value 
(\ref{eq:d2estimate}) estimated 
from the diffusion data.

\begin{figure}
\begin{center}
 \includegraphics[width=10cm,height=8cm]{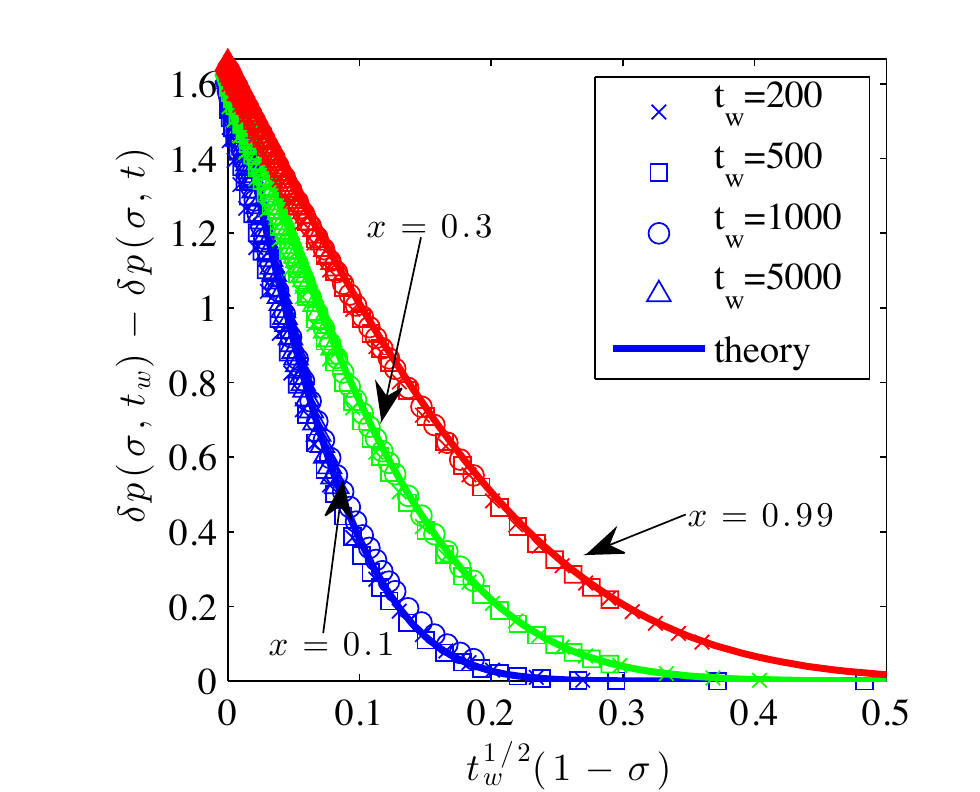}
 \caption{Behaviour of the linear perturbation solution $\delta 
p(\sigma,t)$ in the interior boundary layer at $\sigma=1$. Shown is
$\delta p(\sigma,\tw)-\delta p(\sigma,t)$ versus $\tw^{1/2}(1-\sigma)$ 
for different 
$x=(t-\tw)/\tw$ -- as indicated by the arrows and respective colours -- and 
different $\tw$ as shown by the symbols.
Solid lines give the prediction from the 
scaling theory for $\tw\to\infty$. 
}
\label{fig:interior_profile}
\end{center}
\end{figure}

\begin{figure}
\begin{center}
\includegraphics[width=10cm,height=8cm]{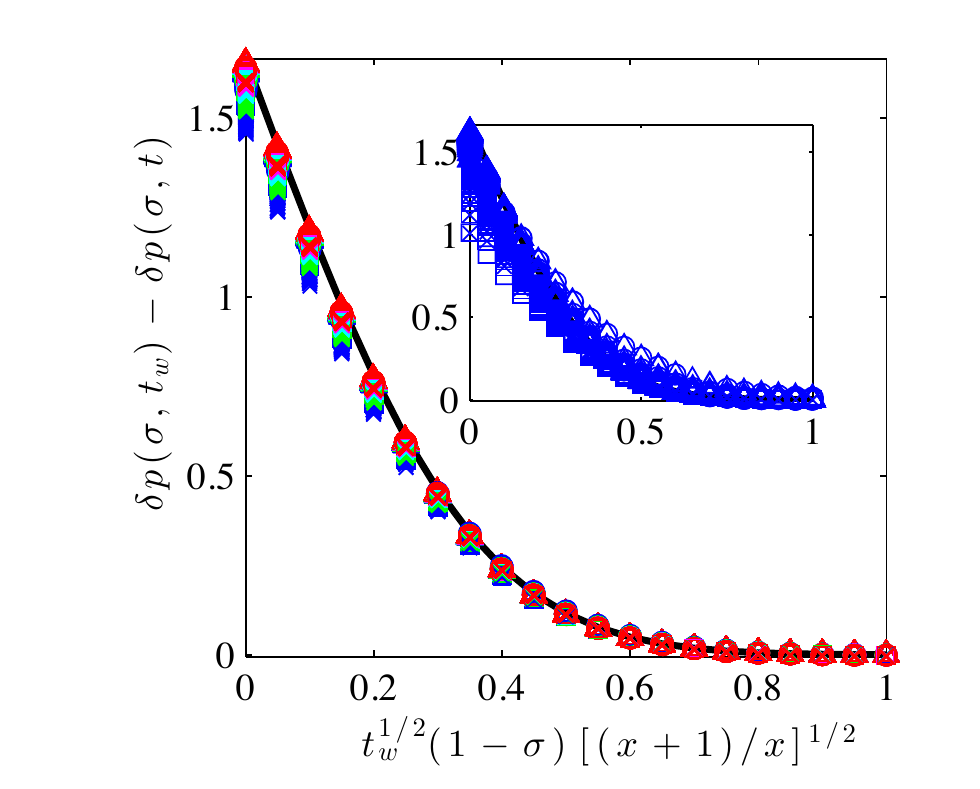}
  \caption{Master curve for the behaviour of the linear perturbation 
solution $\delta 
p(\sigma,t)$ in the interior boundary layer at $\sigma=1$. Shown is
$\delta p(\sigma,t_w)-\delta p(\sigma,t)$ versus 
$\tw^{1/2}(1-\sigma)[(x+1)/x]^{1/2}$ for 
a range of different $x=(t-\tw)/\tw$.
Symbols and colors have the same meaning as 
in Fig.~\ref{fig:fitting_perturbed_tail}. Main plot shows values of 
$x>0.1$ 
  while inset shows values of $x<0.1$.
}
  \label{fig:interior_stacked}
  \end{center}
\end{figure}

Next we consider the behaviour $\delta p(\sigma,t)$ in the interior 
$-1<\sigma<1$ where it is a little more complex. We want to confirm in 
particular that the boundary layer sizes scale as $\tw^{-1/2}$ rather than 
as $\tw^{-1}$ in the exterior.
In the interior there are two boundary layers, around $\sigma=0$, and 
$\sigma=1$ (with a mirror image around $\sigma=-1$. As the scaling is the 
same except for prefactors -- compare (\ref{eq:B0}) 
and 
(\ref{eq:C0}) 
-- we focus on $\sigma=1$.
From the theory we have here to leading order
\begin{equation}
 \delta p(\sigma,t)\approx \delta Q_0(\sigma,0)+\delta 
Q_0(1,0)\left\{-1+\mbox{erf}(z/[2(\tilde{d}_2
x/(1+x))^{1/2}])\right\}.
\end{equation}
where now $z=(1-\sigma)\tw^{1/2}$. The first term, $\delta 
Q_0(\sigma,0)$, depends on the preparation of the system
and so is not known {\em a priori}. But it drops out if we consider 
$\delta p(\sigma,\tw)-\delta 
p(\sigma,t)$. We plot this quantity against $(1-\sigma)\tw^{1/2}$ in 
Fig.~\ref{fig:interior_profile} for three values of $x$ and 
several values of $\tw$, in analogy to Fig.~\ref{fig:perturbed_tail} for 
the exterior boundary layer.
In Fig.~\ref{fig:interior_stacked} we show
the master curve of $\delta p(\sigma,\tw)-\delta p(\sigma,t)$ against 
$(1-\sigma)\tw^{1/2}[(x+1)/x]^{1/2}$ for several values of $x$ and $\tw$. 
Note that since $\delta p(\cdot,\tw)$ is not computed on the same mesh as 
$\delta p(\cdot,t)$, we used linear interpolation to estimate $\delta 
p(\cdot,\tw)$ on the finer mesh.

\section{Summary and outlook}
\label{sec:summary}

We have studied the aging dynamics of the H\'ebraud-Lequeux (HL) model for 
the flow of amorphous materials, and the linear stress response to applied 
shear strain that it produces. Physically, the main qualitative conclusion 
is that the HL model in its glass phase freezes in a manner that depends on 
its initial preparation. This is because the diffusion constant $D(t)$ that 
drives stress relaxation decays so quickly that only a finite amount of 
memory of the initial condition can be erased. Accordingly the two-time 
stress relaxation function $G(t,\tw)$ does not decay fully even for 
$t\to\infty$, and its plateau value increases as the system becomes more 
elastic with in creasing age $\tw$. The explicit result, from 
(\ref{G_intuitive_form}), is $G(t,\tw) = 1 - \mbox{const}\times 
(1/\tw-1/t)^{1/2}$.

The same physics of course drives the response to oscillatory strain as 
measured by the viscoelastic spectrum. In the relevant limit of low 
frequencies $\omega$ and long measurement times $t$ we found that this 
behaves as $G^*(\omega,t) = 1-\mbox{const}\times (1-i)/[t\sqrt{\omega}$. 
While the standard expectation for a system with simple aging would be a 
function depending on $\omega t$ only, here the deviation from purely 
elastic behaviour is not simply $1/\sqrt{\omega t}$, but is suppressed by 
an additional factor of $1/\sqrt{t}$.

Mathematically, what is interesting is that the HL requires rather 
different tools of analysis for its aging dynamics than e.g.\ for models 
such as SGR~\cite{Sollich98}, where one can use temporal Laplace transforms 
and characteristics. In the HL case, most of the physics happens around the 
stress threshold ($=1$) above which yielding occurs. This necessitates the 
use of boundary layer techniques that have previously been deployed by one 
of us to understand the glass transtion in the HL 
model~\cite{Ol,OlRe1,OlRe2}.

We saw that the boundary layer scaling with age $\tw$ is different below 
and above the stress treshold in the linear response $\delta p(\sigma,t)$ 
of the stress distribution to step strain. In effect, yielding above the 
treshold is sufficiently fast for the stress relaxation to become dominated 
by diffusion of stress values $\sigma$ from below to just above the 
threshold, where yielding takes place effectively instantaneously.

Returning finally to the physical implication of our results, it seems to 
us that there is little evidence in experimental data on soft amorphous 
materials for the incompletely decaying stress relaxation $G(t,\tw)$ -- and 
the increasing height of its plateau with age $\tw$ -- that we found for 
the HL-model in its aging phase. As the model has been widely used, also as 
a mean field description for spatially resolved model variants, it is 
important to be aware of this limitation. The insight that the freezing 
behaviour arises from a lack of ``self-sustaining'' noise in the model -- 
as reflected in the rapid decay of $D(t)$ -- may also help to develop more 
sophisticated variants of the model (e.g.~\cite{BouGuaTarZam16}).

\appendix
\section{Aging without strain}  
\label{app:aging_no_strain}

In this Appendix we study the equations for the aging dynamics, (\ref{int}) 
and (\ref{ext}) with boundary conditions (\ref{cont},\ref{contd}). Our aim 
is to find the exponent parameters $l$ and $s$, and from these determine 
the leading behaviour of $p(\sigma,t)$ summarized in 
(\ref{R1sol},\ref{Q1sol}). We show first, via several intermediate steps, 
that $l\leq s \leq 2l-1$. Assuming then a generic initial condition we 
further show that $l=s$, and argue that in fact $l=s=1$. The method will 
centre around determining for which $k$ the exterior functions $R_k^\pm(z)$ 
and the coefficients $d_k$ can be zero or non-zero.

\subsection{General arguments}

Starting with the $d_k$, define $n$ as
\be
n=\min\{k|d_k\neq0\},
\ee 
 Then in the sum in 
(\ref{ext}) the largest value of $k$ that contributes is $m+l-n$:

\be
-\frac{m-s}{s} R^\pm_{m-s}
+ \frac{l}{s}z R^{\pm'}_{m-s}
= \sum_{k=0}^{m+l-n} d_{m-k+l} R^\pm_k{}'' - R^\pm_m
\label{ext2}
\ee

We can deduce from this that $n=l$, proceeding by contradiction.

If $n>l$, then the term of the largest order (containing the $R^\pm_k$
with the largest $k$) is the last one on the r.h.s.\ of (\ref{ext2}). 
Setting $m=0$
gives then $R^\pm_0=0$, and inductively one sees that all $R^\pm_k=0$,
in contradiction to $d_n \neq 0$. Conversely, if $n<l$, then the term
of the largest order is the $k=m+l-n$ contribution from the sum. Setting
$m=n-l$ then gives $d_n R^\pm_0{}''=0$ and hence $R^\pm_0=0$, and
inductively $R^\pm_k=0$, leading again to a contradiction. Having excluded 
both $n>l$ and $n<l$ proves that $n=l$, and so
\be
d_0 = \ldots = d_{l-1}=0.
\label{d_zero}
\ee

Our next step is to deduce that also $R^\pm_k= 0$ for $k=0,\ldots, l-1$, 
and to find the first nonzero function, $R^\pm_l$.
Choosing $m=0$ in
(\ref{ext2}) yields
\be
0=d_l R^\pm_0{}''- R^\pm_0
\ee
Now $d_l$ as the leading term in $D(t)$ has to be positive, hence
(bearing in mind also the definition of $d^\pm_0$) one has $R^\pm_0
=d^\pm_0/(\alpha\sqrt{d_l}) \exp(-z/\sqrt{d_l})$. Since $R^\pm_0$
gives the leading contribution to $p(\sigma,t)$ in the exterior, it has to 
be
non-negative, hence $d^\pm_0\geq 0$. But $d_0=d_0^++d_0^-=0$ (from 
(\ref{d_zero}) together with $l\geq 1$), so
we have $d^\pm_0= 0$ and therefore $R^\pm_0=0$. Repeating the argument, one 
proves inductively
that 
\be
R^\pm_0=\ldots=R^\pm_{l-1}=0
\label{R_zero}
\ee
The first nonzero functions $R^\pm_k$ are
then
\be
R^\pm_l = \frac{d^\pm_l}{\alpha\sqrt{d_l}} e^{-z/\sqrt{d_l}}
\label{R_l}
\ee

Next we look at the interior profiles $Q_k$. We show that $s\leq 2l-1$; 
that the $Q_k$ vanish for $k=1,\ldots, 2l-s-1$; and that $Q_{2l-s}$ has to 
be nonzero. We start from (\ref{int}) and note that because of 
(\ref{d_zero}), the largest $k$ 
that can contribute in the sum is the one where $m-k-l=l$, i.e.\ $k=m-2l$:
\be
-\frac{m-s}{s} Q_{m-s} = \sum_{k=0}^{m-2l} d_{m-k-l} Q''_k +
d_{m-l} \frac{\delta(\sigma)}{\alpha}
\label{int2}
\ee
Now if we had $s\geq 2l$, we could choose $m=2l$ and the l.h.s.\ would be 
zero (either because of the prefactor, for $s=2l$, or because $m-s<0$ so 
that $Q_{m-s}$ vanishes). This would give, after division by $d_l>0$, 
\be
0 = Q''_0 + \frac{\delta(\sigma)}{\alpha}
\label{unsolvable}
\ee
The solution of this equation would consist of two line segments, one for 
each of the regions $-1<\sigma<0$ and $0<\sigma<1$, but it is easy to show 
that in the glassy regime ($\alpha<1/2$) this solution would violate either 
positivity ($Q_0(\sigma)\geq 0$) or normalization ($\int_{-1}^1 d\sigma\ 
Q_0(\sigma)=1$). The assumption $s\geq 2l$ has led to a contradiction, 
hence $s\leq 2l-1$ as announced.

Choosing successively $m=s+1,\ldots, 2l-1$ in (\ref{int2}) then shows that
\be
Q_1=\ldots=Q_{2l-s-1}=0
\label{vanishing_Q}
\ee
provided that $s\leq 2l-2$, otherwise there
are no $m$ in the required range. On the other hand we have shown in 
(\ref{R_l}) that at least one of the functions $R^\pm_l$
has to be nonzero, hence from the boundary 
condition
(\ref{cont}) also $Q_l$ cannot be identically zero. Comparing with 
(\ref{vanishing_Q}) yields 
$l \geq (2l-s-1)+1$, hence $s\geq l$. Overall, we have now deduced that 
$l\leq s \leq 2l-1$.

To get the first non-vanishing interior profile we set $m=2l$ in 
(\ref{int2}) to find
\be
-\frac{2l-s}{s}Q_{2l-s} = d_{l} Q''_0 + d_l \frac{\delta(\sigma)}{\alpha}
\label{Q2ls}
\ee
$Q_{2l-s}$ cannot vanish identically as otherwise we would get back to our 
previous equation (\ref{unsolvable}) for $Q_0$ that has no valid solution.
Instead we can use (\ref{Q2ls}) to determine $Q_{2l-s}$ from $Q_0$, and 
then recursively $Q_{2l-s+1}$,
$Q_{2l-s+2},\ldots$ for
$m=2l+1,2l+2,\ldots$ What is notable is that we never
get a closed equation for $Q_0$, which means that this profile must
depend on the initial conditions. 

\subsection{Generic initial conditions lead to $l=s$}

Consider now the generic case where one expects that the second derivatives
$Q''_0(\pm1)$ will not both vanish. Then one of $Q_{2l-s}(\pm1)$ is
nonzero from (\ref{Q2ls}), and from (\ref{cont}) also the corresponding 
$R^\pm_{2l-s}(0)$ cannot vanish identically. But we know from 
(\ref{R_zero}) that all
$R^\pm_k$ up to $k=l-1$ vanish, so $2l-s\geq l$, i.e.\ $l\geq s$. Together 
with $l\leq s$ as shown in the previous subsection, we therefore have in 
the generic case $l=s$. 

One can show further that the $Q_k$ and $R^\pm_k$ vanish when $k$ is not a 
multiple of 
$s$, so that without loss of generality one can take $l=s=1$. 
These values imply that the boundary layer has width $1/t$, and the 
diffusion constant scales as $D\sim t^{-2}$ to leading order. This is 
consistent with initial condition-dependent freezing, because $\int D(t) 
dt<\infty$.

With $l=s=1$, the leading order equations are now 
\bea
R^\pm_1(z) &=& \frac{d^\pm_1}{\alpha\sqrt{d_1}} e^{-z/\sqrt{d_1}} 
\label{Rsol}\\
-Q_{1}(\sigma) &=& d_{1} Q''_0(\sigma) + d_1 \frac{\delta(\sigma)}{\alpha}
\label{Qsol}\\
R^\pm_1(0) &=& \frac{d^\pm_1}{\alpha\sqrt{d_1}}\ =\
Q_1(\pm 1)\ =\ -d_1 Q''_0(\pm 1) \label{Qsec}\\
R^\pm_1{}'(0) &=& -\frac{d^\pm_1}{\alpha d_1}\ =\ \pm Q_0'(\pm1) 
\label{Qprem}
\eea
Recalling that $d_1= d_1^+ + d_1^-$, the last line (\ref{Qprem}) gives
\be
-Q_0'(1)+Q_0'(-1)=1/\alpha
\label{Q0constraint}
\ee
This is a condition that the frozen profile needs to satisfy: here we have 
an aspect of $Q_0(\sigma)$ that is controlled by the aging dynamics with 
its partial loss of memory, rather than frozen-in initial information.

The result (\ref{Qsec}) shows that the frozen profile also has to
have non-positive $Q_0''(\pm1)$. Quantitatively, starting from 
(\ref{Qsec}), dividing by
$\sqrt{d_1}$ and adding the two cases one has
$-\sqrt{d_1}[Q_0''(1)+Q_0''(-1)]=1/\alpha$ and so
\be
d_1=(-\alpha[Q_0''(1)+Q_0''(-1)])^{-2}
\ee
From (\ref{Qsec}) one then finds in more detail
\be
d^\pm_1 = \frac{-\alpha Q_0''(\pm1)}{(-\alpha[Q_0''(1)+Q_0''(-1)])^{3}}.
\ee
These explicit expressions together with (\ref{Rsol},\ref{Qsol}) 
demonstrate that up to order $1/t$ ($k=1$), all profiles $Q_1$, $R^\pm_1$ 
are fully determined by $Q_0''$.

\subsection{Non-generic case}

We comment briefly on the non-generic case where $Q_0''(\pm1)=0$. 
 Then $Q_{2l-s}(\sigma)$, which we already know 
does not vanish identically, is nonetheless zero for
$\sigma=\pm1$. The boundary condition
(\ref{cont}) now implies that $R^\pm_{2l-s}(0)=0$, whereas at least one of 
$R^+_l(0)$ and $R^-_l(0)$ is nonzero from (\ref{R_l}). Hence $l\geq 2l-s+1$ 
or $s\geq l+1$,
which means that the width $\sim t^{-l/s}$ of the boundary layer is {\em 
larger} than in the generic case, decaying more slowly with time. We 
already know that $l\leq s\leq 2l-1$, so the smallest $l$ possible in the 
non-generic case is $l=2$, which would then imply $s=3$.

One can now ask about the next order in $Q_k$ beyond $k=2l-s$. Choosing 
$m=2l+1$ in (\ref{int2}) gives
\be
-\frac{2l+1-s}{s} Q_{2l+1-s} = d_{l+1} Q''_0 + d_l Q''_1 + 
d_{l+1} \frac{\delta(\sigma)}{\alpha}
\label{non-generic}
\ee
If $s=2l-1$, which is true for e.g.\ the choice $(l=2,s=3)$, then $Q_1$ 
obeys
\be
-\frac{1}{s} Q_1 = d_l Q_0''+d_l\frac{\delta(\sigma)}{\alpha}
\ee
from (\ref{Q2ls}) and does not vanish identically. Inserting into
(\ref{non-generic}) shows that $Q_{2l+1-s}\equiv Q_2$ then behaves as
$Q''_2(\pm1)\propto Q_1''(\pm1)\propto Q_0^{(4)}(\pm1)$ at the
boundary. This suggests a further case division depending on whether
these fourth derivatives both vanish, and there is likely to be a
hierarchy of such further divisions depending on derivatives of
increasing order of $Q_0(\sigma)$ at the boundaries. If at least one
of the fourth derivatives is nonzero, then one of $Q_2(\pm 1)$ is also
nonzero, hence from (\ref{cont}) so is one of $R_2^\pm(0)$. This
implies $l\leq 2$, hence in fact $l=2$ because as shown above smaller
$l$ are impossible in the non-generic case.

Above we had assumed $s=2l-1$, and we would conjecture that the
opposite case $s\leq 2l-2$ can be excluded using similar arguments as
in generic case. We will not explore this issue further here, however,
as the non-generic case is unlikely to be relevant for physically
plausibe initial conditions.

\section{Calculations for complex shear modulus}
\label{app:G_omega}

We first show that in the HL model the waiting-time dependent oscillatory 
shear modulus and the forward shear modulus become identical in the 
long-time limit.

To find an expression for the full $\tw$-dependent shear modulus 
(\ref{eq:fullGstar}), we insert the expression (\ref{eq:Gasymp}) for the 
stress relaxation function $G(t,\tw)$ in the long time regime.
We set $w=\omega t$, 
$w'=\omega(t-t')$ and recall that $x=(t-\tw)/\tw$. One obtains with a few 
lines of algebra
\begin{equation}
 G^*(\omega,t,\tw)=1-\frac{c}{\sqrt{t}}\left(
\sqrt{x} \eee^{-\ii w(1-r)} + \ii \int_0^{wx/(1+x)} dw'\, 
\sqrt{\frac{w'}{w-w'}} \eee^{-\ii w'}
\right)
\end{equation}
The term in brackets can be written as
\be
\ii \int dw' f_{w,x}(w') \eee^{-\ii w'}
\ee
where
\begin{eqnarray}
f_{w,x}(w') &=\sqrt{\frac{w'}{w-w'}} \qquad &\mbox{for} \quad w'<wx/(1+x)\\
f_{w,x}(w') &= \sqrt{x} \qquad &\mbox{for} \quad w'>wx/(1+x) 
\end{eqnarray}
On the other hand the forward spectrum can be written in the long-time 
limit as (\ref{forward_spectrum_long_time}):
\begin{equation}
G^*(\omega,t) = 1- \frac{c}{\sqrt{t}} \,\ii \int_0^{\infty} dw' 
\sqrt{\frac{w'}{w+w'}} \eee^{-\ii w'}
\label{forward_spectrum_again}
\end{equation}
One now sees that for any fixed $w'$, the two integrands become identical, 
both approaching $\sqrt{w'/w}$, provided that $w$ and $wx/(1+x)$ are large 
(and in particular larger than $w'$). So in this limit the full 
$\tw$-dependent 
spectrum and the forward spectrum do indeed become identical.

Let us now compute the forward spectrum by computing 
the integral in (\ref{forward_spectrum_again}). For the intermediate 
calculations it is convenient to express the dependence on $w$ in terms of 
$z=w/2$.
We perform two changes of variables to carry out the integration over 
$w'$. First we put 
$z'=w'/z+1$. This leads to the equality
\begin{eqnarray}
 \int_{0}^\infty \dd 
w'\, \eee^{-\ii w'}\sqrt{\frac{w'}{w+w'}}&
=z\eee^{\ii z}\int_1^{\infty}
\dd z'\,
\eee^{-\ii zz'}\sqrt{\frac{z'-1}{z'+1}}
\label{aux_integral}
\end{eqnarray}
Now set $z'=\cosh(t)$ and use the hyperbolic trigonometric relations 
$\cosh(t)-1=2\sinh^2(t/2)$, $\cosh(t)+1=2\cosh^2(t/2)$ and
\begin{eqnarray}
 \dd z'&=&\sinh(t)\dd t=2\sinh(t/2)\cosh(t/2)\dd t
\end{eqnarray}
Thus we find for our integral
\begin{eqnarray}
&=z \eee^{\ii z}\int_0^{\infty}
\dd t\,\eee^{-\ii 
z\cosh(t)}\sqrt{\frac{2\sinh^2(t/2)}{2\cosh^2(t/2)}}
2\sinh(t/2)\cosh(t/2)\\
&=2z \eee^{\ii z}\int_0^{\infty}\dd t\,\eee^{-\ii z\cosh(t)}\sinh^2(t/2)\\
&=z \eee^{\ii z}\int_0^{\infty}\dd t\,\eee^{-\ii z\cosh(t)}(\cosh(t)-1)
\end{eqnarray}
The remaining integral can be expressed in terms of Hankel functions~\cite{AbSt} defined as:
\begin{equation}
 H_\nu^{(2)}(z)=-2\frac{\eee^{\ii
\nu\pi/2}}{\pi}\int_0^{\infty}\eee^{-\ii z\cosh(t)}\cosh(\nu t)\dd t
\end{equation}
This gives eventually
\begin{equation}
G^*(w,t)=1+\frac{c\pi}{4\sqrt{t}} w\eee^{\ii w/2}\left(H_0^{(2)}
(w/2)+\ii H_1^{(2)}(w/2)\right)
\end{equation}
From asymptotic properties of the Hankel functions, one can then obtain for $w\gg 1$ the expression (\ref{G_omega_final}) in the main text.

\bibliographystyle{unsrt}
\bibliography{HL-JPhys-A-v1,references,HL-refs}
\end{document}